\documentclass[amsmath,twocolumn,showpacs,aps,pra]{revtex4}
\usepackage{bm}
\usepackage{graphicx}
\begin{document}

\title{Fluctuations of radiation from a chaotic laser below threshold}
\author{E. G. Mishchenko}
\affiliation{Lyman Laboratory, Department of Physics, Harvard
University, MA 02138} \affiliation{L.D. Landau Institute for
Theoretical Physics, Moscow 117334, Russia}

\begin{abstract}
Radiation from a chaotic cavity  filled with  gain medium is
considered. A set of coupled equations describing the photon
density and the population of gain medium is proposed and solved.
The spectral distribution and fluctuations of the radiation are
found. The full noise is a result of a competition between
positive correlations of photons with equal frequencies (due to
stimulated emission and chaotic scattering) which increase
fluctuations, and a suppression due to interaction with a gain
medium which leads to negative correlations between photons. The
latter effect is responsible for a pronounced suppression of the
photonic noise as compared to the linear theory predictions.
\end{abstract}
\pacs{42.50.Ar, 05.40.-a, 42.68.Ay} \maketitle

{\it Introduction.} Amplifying random media have now been a focus
of intense research for almost a decade
\cite{LBG,SLA,WAL,ZCY,GCN}. The ultimate goal of these efforts is
a construction of a random laser in which laser action utilizes
the feedback provided by multiple scattering from disorder. Lasing
action was reported in a recent work by Cao {\it et al.}
\cite{Cao,Cao2} and Frolov {\it et al.} \cite{Frol}. The
difference between the true coherent random laser (RL) and the
amplified spontaneous emission (ASE) cannot be identified by the
emitted light intensity or its frequency profile as  both ASE and
RL are characterized by a pronounced narrowing of the spectral
linewidth $\gamma $ compared to the atomic transition width
$\Gamma$. The knowledge of the statistics of light fluctuations is
therefore necessary to make a distinction. While RL action should
be coherent and therefore obey the Poissonian distribution, the
ASE fluctuations are  above the Poissonian level (super-Poissonian
noise). Both cases seem to be realized in experiment, that is of
RL \cite{Cao2,PCV} and ASE \cite{ZPF}. There is still no complete
theory able to predict which case ought to be expected in a
particular system. From the outset, one expects RL regime to occur
whenever there is a single mode above the laser threshold or a few
non-overlapping modes, and ASE case to be realized as long as
there is a number of overlapping modes above the threshold. In the
latter case the scattering between the lasing modes should lead to
the enhancement of fluctuations. Since the broadening of the laser
modes is given roughly  by the inverse mean dwell time
$\tau^{-1}_{dw}$, cavity modes overlap strongly provided that the
mean level spacing $\Delta$ is small, $\tau^{-1}_{dw} \gg \Delta$.
(Hereinafter, we use the units with $\hbar=1$.)

 Although there is by now a substantial literature on various
models for random lasers \cite{XS,VS,BRC,ARS,D}, we address only
those papers that consider fluctuations of laser radiation.
Beenakker put forward a quantum approach for a photon statistics
based on the random matrix theory \cite{Ben} for a linear
amplifier below laser threshold. The threshold is achieved when
the dwell time inside the system $\tau_{dw}$ exceeds the
characteristic amplification time determining the rate for photon
emission $\tau_a$. The threshold features a singularity in the
photon density. The fluctuations of radiation demonstrate even
stronger singularity as the ratio of the noise power to the mean
photon flux $S/\overline{J}$ diverge at the threshold, as a result
of scattering between many lasing modes. Linear theory fails upon
approaching the threshold when the assumption of equilibrium gain
breaks down. Hackenbroich {\it et al.} \cite{HVE} considered
statistics of a single lasing mode in a chaotic cavity above the
threshold. Cavity opening was assumed to be covered with a barely
transparent mirror making the inverse dwell time small compared to
the spacing between the modes, $\tau_{dw}^{-1} \ll \Delta$, thus
preventing them from overlapping. In Ref.\  \cite{P} the nonlinear
effects have been numerically studied for a system with an opening
much smaller than the wavelength. While such a model might be
useful for a microwave emission, its relevance to random lasers
with typical openings much wider than the wavelength of light
remains unclear.

\begin{figure}[h]
\resizebox{.24\textwidth}{!}{\includegraphics{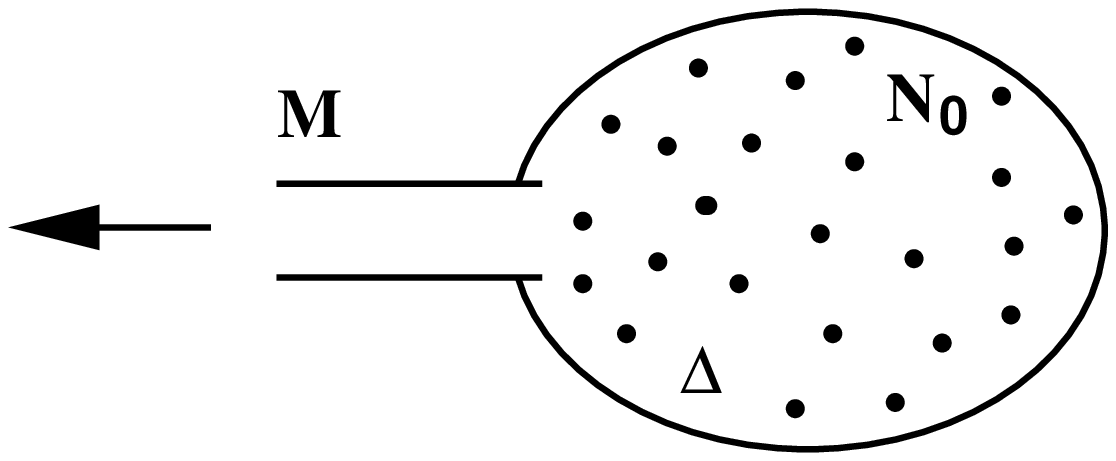}}
\caption{Chaotic laser: a cavity characterized by the mean level
spacing $\Delta$ contains $N_0$ particles of a gain medium and is
connected to the outside via a waveguide (lead) supporting $M$
transverse channels.}
\end{figure}

{\it Chaotic Laser.} In the present paper we develop a
semiclassical approach based on the radiative transfer theory
\cite{MB} for a chaotic cavity filled with a gain medium, see
Fig.\ 1 (let us, following Refs.\ \cite{HVE} and \cite{D}, call
such a system 'chaotic laser'). The cavity is coupled to the
outside through a waveguide supporting $M \gg 1$ transverse modes,
so the inverse dwell time $\tau^{-1}_{dw}=\frac{M\Delta}{2\pi}$ is
large compared to the mean level spacing $\Delta =
\frac{\pi^2c^3}{\omega^2L^3}$, which is inversely proportional to
the photon density of states and a system volume $L^3$. The
ergodic time $\tau_{er}=L/c$ is the shortest time scale of the
system, $\tau_{er} \ll \tau_{dw}, \tau_a$. This condition assures
that photons scatter chaotically from the boundaries many times
before leaving the cavity or stimulating an emission of another
photon, therefore resulting in the homogeneous and isotropic
distribution $f_\omega$ everywhere inside the cavity. The number
of photons inside the cavity with the frequency in the interval
$d\omega$ is given by $f_\omega d\omega/\Delta$. The rate at which
this distribution changes in time,
\begin{equation}
\label{photon_lin} \frac{\partial f_{\omega}}{\partial  t}+
J_\omega \Delta = I_\omega,
\end{equation}
is determined by the photon flux via the waveguide, $J_\omega$ and
by the photon creation and absorption inside the cavity
$I_\omega$. The flow of photons through the waveguide (lead)  is
proportional to the size of the opening (characterized by the
number of channels $M$) and the difference between the
concentration of photons inside and outside the cavity, the latter
denoted by $n_\omega$ (in general case of external radiation
incident on the system),
\begin{equation}
\label{current}
J_\omega=\frac{M}{2\pi}~[f_{\omega}-n_\omega]+{\cal J}_\omega,
\end{equation}
with the last term standing for the Langevin source describing
fluctuations in the system. The left-hand side of the rate
equation (\ref{photon_lin}) has a familiar form of the particle
conservation condition while the right-hand side accounts for the
photon creation and absorption inside the cavity,
\begin{equation}
\label{emission} I_\omega= w_\omega^+ (f_\omega +1) -w_\omega^-
f_\omega +{\cal L}_\omega \Delta.
\end{equation}
The quantities $w_\omega^+$ and $w_\omega^-$ stand for the rates
of photon emission and absorption respectively. The probability of
photon creation $w_\omega^+ (f_\omega +1)$ takes into account both
stimulated and spontaneous emission as required by the quantum
statistics of photons.

In addition, Eqs.\ (\ref{photon_lin}-\ref{emission}) contain
stochastic Langevin sources   which are necessary to take into
account the randomness of photon transmission through  the lead
${\cal J}_\omega$, and the randomness of the  emission and
absorption events, ${\cal L}_\omega$. These terms have zero
average, $\overline{{\cal J}}_\omega=\overline{{\cal
L}_\omega}=0$, and correlators which follow from the assumption
that elementary stochastic events of scattering, emission and
absorption have independent (and therefore Poissonian)
distributions. In particular, the Langevin term associated with
non-conservation of particles, ${\cal L}_\omega={\cal
L}^+_\omega+{\cal L}^-_\omega$ consists of separate contributions
from emission and absorption. Different processes have independent
distributions (in particular, they are uncorrelated when occur at
different times) with the second moments given simply by their
average rates (property characteristic for Poissonian processes),
\begin{eqnarray*}
\overline{{\cal L}^+_\omega (t){\cal L}^+_{\omega'}
(t')}&=&\delta_{\omega t}
{\Delta^{-1}}w_\omega^+(\overline{f}_\omega+1),\\
\overline{{\cal L}^-_\omega (t){\cal L}^-_{\omega'}
(t')}&=&\delta_{\omega t} {\Delta^{-1}} w_\omega^-
\overline{f}_\omega, ~~ \overline{{\cal L}^+_\omega (t){\cal
L}^-_{\omega'} (t')}=0,
\end{eqnarray*}
where $\delta_{\omega t}$ is the shorthand for $\delta
(\omega-\omega')\delta (t-t')$. Using these formulas we arrive to
the following expression,
\begin{equation}
\label{lan11} \overline{{\cal L}_\omega (t){\cal L}_{\omega'}
(t')}=\delta_{\omega t}
{\Delta^{-1}}[w_\omega^+(\overline{f}_\omega+1) +w_\omega^-
\overline{f}_\omega].
\end{equation}
Similar reasoning allows us to write the correlation function for
the Langevin sources describing randomness in the photon flux
through the waveguide,
\begin{equation}
\label{lan12} \overline{{\cal J}_\omega (t){\cal J}_{\omega'}
(t')}=\delta_{\omega t}
\frac{M}{2\pi}[\overline{n}_\omega(\overline{n}_\omega+1)
+\overline{f}_\omega(\overline{f}_\omega+1) ].
\end{equation}
Since we assume no reflection in the lead, the only remaining
source of fluctuations is the randomness of the photon densities
at both ends of the waveguide \cite{refl}.

 In equilibrium
at temperature $T$, the  function $f_\omega$ must be equal to the
Bose-Einstein distribution. This fixes the ratio,
$w_\omega^-/w_\omega^+= e^{\omega/T}$. So far our discussion has
been completely general, valid for both the absorbing, $w_\omega^-
> w_\omega^+$ ($T>0$), and amplifying $w_\omega^-
< w_\omega^+$ ($T<0$), media. Hereinafter we concentrate for
simplicity on the case of complete population inversion,
$w_\omega^-=0$ ($T=0^-$). The complete population inversion is
realized, e.g.\ when the radiative transition occurs not to a
ground state but rather to some excited state with very short
(non-radiative) lifetime, see Fig.\ 2.

\begin{figure}[h]
\resizebox{.45\textwidth}{!}{\includegraphics{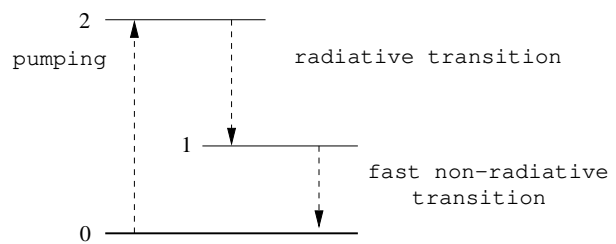}}
\caption{Complete population inversion realized in a three-level
laser. The atom of the gain medium is pumped to an excited state
2, the photon is emitted when the atom undergoes a transition to a
lower excited state 1. If the relaxation from the excited state 1
to the ground state 0 is fast, the excited state 1 is almost
always empty. Therefore the radiative transition 2$\to$1 occurs
with the effective temperature $T=0^-$.}
\end{figure}
We assume  the spectral line for atomic transitions to be a
Lorentzian  centered around the frequency $\omega_0$ with the
linewidth $\Gamma$,
\begin{equation}
\label{photons} w^{+}_\omega =a_\omega  N\Delta, ~~~
a_\omega=\frac{w_0 \Gamma }{\pi[(\omega-\omega_0)^2+\Gamma^2]}
\end{equation}
where $N$ is the number of excited atoms of the gain medium in the
cavity, and $w_0$ is the total probability (per unit time) for an
excited atom to spontaneously emit a photon into vacuum.

The dynamics of the gain medium follows the equation,
\begin{equation}
\label{gain} \frac{\partial N}{\partial  t}= P (N_0-N)
-\frac{N}{\tau} -\Delta^{-1}   \int d\omega ~ I_\omega +{\cal N},
\end{equation}
where $P$ is the pumping rate, $N_0$ is the total number of gain
medium particles, and $\tau^{-1}$ is the probability for a
non-radiative relaxation. The third term is the total loss from
radiative transitions into all possible cavity modes. Note that it
contains stochastic emission source, as specified by Eq.\
(\ref{emission}). The Langevin term ${\cal N}$ describes the
spontaneous fluctuations in the number of excited atoms, which are
also assumed Poissonian,
\begin{equation}
\label{lan3} \overline{{\cal N}(t){\cal N}(t')}=\delta (t-t')
[P(N_0-\overline{N})+{\overline{N}}/{\tau} ].
\end{equation}
Note that different Langevin terms ${\cal J}_\omega$, ${\cal
L}_\omega$, ${\cal N}$ are independent and have zero
cross-correlators. The coupled equations (\ref{photon_lin}),
(\ref{gain}) together with the correlators
(\ref{lan11}-\ref{lan12}), (\ref{lan3}) determine the radiation
from a chaotic cavity.

{\it Mean photon distribution.} We illustrate the above formalism
by calculating the average photon flux from the cavity and its
fluctuations assuming no radiation incident on the system,
$n_\omega=0$. In the stationary regime the time-averaged equation
for the gain medium (\ref{gain}) is easily solved and  the number
of excited atoms is expressed via the photon distribution,
\begin{equation}
\label{n} \overline{N}=\frac{PN_0}{P_1+w_0+\int d \omega ~
a_\omega \overline{f}_\omega},
\end{equation}
where $P_1=P+\tau^{-1}$. Substituting it into Eqs.\
(\ref{photon_lin}-\ref{emission}) one obtains an integral equation
for the photon distribution function, which could be solved
analitically. The time-averaged distribution has the form of a
Lorentzian,
\begin{equation}
\label{phot} \overline{f}_{\omega}=
\frac{\Gamma^2-\gamma^2}{(\omega-\omega_0)^2+\gamma^2},
\end{equation}
with the linewidth $\gamma$ determined from the algebraic
equation,
\begin{equation}
\label{width} \Gamma^2-\gamma^2=\frac{2N_0w_0\Gamma\gamma P}{M(
w_0 {\Gamma}+ P_1\gamma)}.
\end{equation}
The total photon flux leaving the cavity
$\overline{J}=\frac{M}{2\pi} \int d\omega \overline{f}_\omega$ is
found from Eqs.\ (\ref{phot}-\ref{width}),
\begin{equation}
\overline{J}=\frac{N_0w_0\Gamma P}{w_0\Gamma +P_1\gamma}.
\end{equation}
The cubic equation (\ref{width}) allows easy numerical solution
for any values of the pumping strength $P$ and the number of
channels $M$. Here we analyze in detail the more illuminating case
of absent non-radiative losses, $\tau^{-1}=0$.

It is convenient to introduce the dimensionless pumping strength
and the effective number of radiative transitions,
$$
p=\frac{P}{w_0}, ~~~\eta=\frac{2N_0w_0}{M\Gamma}.
$$
The parameter $\eta$ is the ratio of the photon dwell time in the
cavity, $\tau_{dw}=2\pi/M\Delta$, and the characteristic time of
spontaneous photon emission, which at the center of the line is
given, according to Eq.\ (\ref{photons}), by
$\tau_a=\pi\Gamma/N_0w_0\Delta$ (if all excited states are
filled). The dimensionless width $g=\gamma/\Gamma$ satisfies a
simple equation,
\begin{equation}
\label{cubic} 1-g^2=\frac{\eta pg}{1+pg}.
\end{equation}
This equation determines the radiation linewidth $g_\eta (p)$ as a
function of the pumping strength for different system geometries:
with the lower values of $\eta$ characterizing open cavities
(large number of outgoing channels $M$) and the higher values of
$\eta$ corresponding to pinching off the waveguide opening. The
other two relevant quantities, the population of the gain medium
and the outgoing flux can also be expressed in terms of the new
dimensionless variables,
\begin{equation}
\frac{\overline{N}}{N_0}=\frac{pg}{1+pg}, ~~~ \frac{2 \overline
J}{M\Gamma} =\frac{\eta p}{1+pg}.
\end{equation}
For weak pumping, $p\ll 1$, the width is narrowed linearly with
the pumping strength,
\begin{equation}
\label{weak} g_\eta(p)=1-\frac{\eta p}{2},
\end{equation}
with the narrowing pronounced stronger for less open cavities
(larger values of $\eta$). The strong pumping, $p \gg 1$, behavior
is qualitatively different for different values of the effective
number of transitions $\eta$.

(i) Open system, $\eta < 1$ (large number of channels supported by
the waveguide). The linewidth remains finite for any value of the
pumping strength, with,
$$
\lim_{p\to \infty} g_\eta(p) = \sqrt{1-\eta}.
$$
The population of excited atoms becomes saturated,
$\overline{N}\to N_0$, while the outgoing flux reaches a constant
value $\overline{J} \propto (1-\eta)^{-1/2}$. The flux diverges
upon pinching off the waveguide opening and approaching the point
$\eta=1$. This point is referred to as the 'laser threshold' in
the linear amplifier theory \cite{Ben}. However, it is not a real
laser threshold (in a sense discussed towards the end of the
paper) but rather a point where nonlinear effects can no longer be
disregarded. We call it a 'critical point' here.
\begin{center}
\begin{table}[h]
\caption{The dependence of the dimensionless linewidth
$g=\gamma/\Gamma$, population of the gain medium, and the outgoing
photon flux for large values of the pumping intensity $p=P/w_0 \gg
1$, for different values of the effective number of optic
transitions $\eta$ (in the absense of non-radiative losses).}
\begin{tabular}{| c | c | c | c |}
\hline & & & \\
$~~\displaystyle \eta=\frac{2N_0 w_0}{M\Gamma}~~$ & $g_\eta (p)$ &
$\displaystyle\frac{ \overline{N}}{N_0}$ & $\displaystyle\frac{2\overline{J}}{M\Gamma}$\\
 & & & \\
\hline
& & & \\
$\eta<1$ & $\displaystyle \sqrt{1-\eta}$ & 1 & $\displaystyle\frac{\eta}{\sqrt{1-\eta}}$\\
& & & \\
\hline
& & & \\
$\eta=1$ & $\displaystyle \frac{1}{p^{1/3}}$ & 1 & $\displaystyle {p^{1/3}}$\\
& & & \\
\hline
& & & \\
$\eta>1$ & $~~\displaystyle \frac{1}{(\eta-1)p}~~$ &
$~~~~\displaystyle \frac{1}{\eta}~~~~$ &
$~~\displaystyle {\left(\eta -{1}\right)p}~~$\\
& & & \\
\hline
\end{tabular}
\end{table}
\end{center}

(ii) In the critical point, $\eta = 1$, the gain population
remains full at large pumping rate and the outgoing flux diverges:
${\overline J} \propto p^{1/3}$. The linewidth is a slowly
decreasing function of pumping $g \propto p^{-1/3}$.

 (iii) Closed system, $\eta > 1$. In this case
the linewidth decreases inversely proportional to the pumping
intensity. The number of photons leaving the cavity grows linearly
with the pumping. The gain population remains partially empty:
${\overline N}/N_0 \to 1/\eta <1$.
 The results for the strong pumping regime
are summarized in Table 1. The quantitative behavior of the
linewidth is illustrated in Fig.\ 3.

{\it Fluctuations}. We now consider the fluctuations of radiation
around the mean solution obtained above. Both the photon
distribution and the number of excited atoms deviate from their
mean values, $f_\omega=\overline{f}_\omega+\delta f_\omega$,
$N=\overline{N}+\delta N$. The fluctuation of the number of
excited atoms is found from the linearized equation (\ref{gain}),
\begin{equation}
\delta N =\overline{N}~\frac{{\cal N}-\int d\omega{\cal
L}_\omega-\overline{N} \int d\omega ~ a_\omega \delta
f_\omega}{PN_0}.
\end{equation}
\begin{figure}[h]
\includegraphics[width=6.3cm]{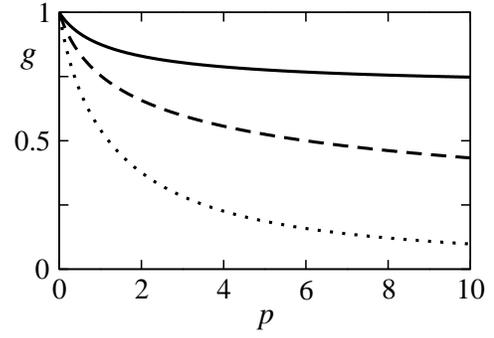}
\caption{The linewidth of radiation from a chaotic laser,
$g=\gamma/\Gamma$, versus the dimensionless pumping intensity
$p=P/w_0$ (neglecting non-radiative losses, $\tau^{-1}=0$) for
different values of the effective number of radiative transitions,
$\eta={2N_0 w_0}/{M\Gamma}$. Solid line: $\eta= 1/2$, the
linewidth $g$ tends to the limit $1/\sqrt{2} \approx 0.71$ for $p
\to \infty$. Dashed line: $\eta=1$, the linewidth decreases as
$g\approx p^{-1/3}$. Dotted line: $\eta =2$, the linewidth behaves
as $g \approx p^{-1}$. }
\end{figure}
Substituting it into the equation (\ref{photon_lin}) we obtain the
linear integral equation for the fluctuation of the distribution
function of photons,
\begin{widetext}
\begin{equation}
\delta f_\omega+\frac{M \overline{f}^{~2}_\omega}{2\pi a_\omega
PN_0} \int d\omega a_\omega \delta f_\omega
=\frac{\overline{f}_\omega({\cal L}_\omega-{\cal
J}_\omega)}{a_\omega \overline{N}} +
\frac{\overline{f}_\omega(\overline{f}_\omega+1)}{PN_0}\left(
{\cal N} -\int d\omega ~{\cal L}_\omega \right)
\end{equation}
Solving this integral equation is rather straightforward and
yields the fluctuation of the outgoing flux,
\begin{equation}
\delta J_\omega = \frac{M}{2\pi}~\delta f_\omega+{\cal J}_\omega=
(\overline{f}_\omega+1) {\cal L}_\omega-\overline{f}_\omega {\cal
J}_\omega- A_\omega \int d\omega' \Bigl[(\overline{f}_{\omega'}+1)
{\cal L}_{\omega'}-\overline{f}_{\omega'} {\cal J}_{\omega'}\Bigr]
+ A_\omega {\cal N}.
\end{equation}
\end{widetext}
In the last expression we utilized the following notations,
\begin{equation}
A_\omega=\frac{M \overline{f}_\omega(\overline{f}_\omega+1)}{2\pi
PN_0 D},~~D=1+\frac{M}{2\pi PN_0}\int d\omega
\overline{f}_\omega^{~2}.
\end{equation}
By using the correlation functions of the Langevin sources
(\ref{lan11})-(\ref{lan12}) and (\ref{lan3}) the
frequency-resolved noise power, $S_{\omega\omega'} =\int dt ~
\overline{{J}_\omega(t) {J}_{\omega'}(0)}$, is found,
\begin{eqnarray}
\label{resolved} S_{\omega\omega'}&=&\delta(\omega-\omega')
\widetilde S_\omega-(A_\omega \widetilde S_{\omega'}+A_{\omega'}
\widetilde S_{\omega})\nonumber\\&&+A_\omega
A_{\omega'}[\widetilde
S+P(N_0-\overline{N})+\tau^{-1}\overline{N}],
\end{eqnarray}
where we introduced the  notations,
\begin{equation}
\label{somega} \widetilde S_{\omega}= \overline{J}_\omega
(\overline{f}_\omega+1)(\overline{f}_\omega^{~2}+\overline{f}_\omega+1),
~~~ \widetilde S=\int d\omega \widetilde S_{\omega}.
\end{equation}
The first term in Eq.\ (\ref{resolved}) represents the
correlations between the photons of the same frequency. Those
correlations are the result of spontaneous and stimulated emission
and also due to chaotic photon scattering inside the cavity.
Assuming again the absence of non-radiative losses we can express
the Fano factor (the ratio of the noise power to the mean flux) at
the center of the line, $\omega=\omega_0$, via a single parameter,
\begin{equation}
\label{fomega} F_{0} =
\frac{S_{\omega_0}}{\overline{J}_{\omega_0}}=
\frac{1-g^2+g^4}{g^6},
\end{equation}
 namely the dimensionless linewidth
$g=\gamma/\Gamma$. This is a monotonously increasing function of
the pumping strength $p$. As before, we analyze three possible
regimes, making use of the above results summarized in Table 1.

(i) Open system, $\eta<1$.  When the pumping intensity increases,
$p \to \infty$, the Fano factor approaches the finite value,
\begin{equation}
\label{fomega1}
 F_{0} = \frac{1-\eta +\eta^2}{(1-\eta)^3}.
\end{equation}
This expression reproduces the result of the random matrix theory
\cite{Ben} for a linear amplifier with a complete population
inversion. Indeed, in an open system, $\eta< 1$, the gain medium
dynamics becomes suppressed for large values of the pumping
strength when all excited states become occupied.

(ii) Critical point, $\eta =1$. For large pumping intensity, $p\gg
1$, the Fano factor grows quadratically,
$$
F_{0} = p^2.
$$

(i) Finally, in a closed system, $\eta>1$, the Fano factor behaves
as the sixth power of the pumping strength $p$ for $p \gg 1$,
$$
F_{0} = (\eta-1) p^6.
$$
The quantitative behavior of the Fano factor (\ref{fomega}) is
illustrated in Fig.\ 4 where numerical results are presented.

{\it Cross-correlations.} Contrary to the linear theory predicting
the photons of different frequencies to be uncorrelated, we
observe significant off-diagonal correlations, as attributed by
the expression (\ref{resolved}). This is due to the strong
coupling of photons to the fluctuations of the gain medium. The
important fact is that these correlations are {\it negative}. It
could be understood from the following reasoning: the emission of
a photon at some frequency reduces the population of excited atoms
making therefore the emission of another photon less likely.

To illustrate better the role of cross-correlations it is useful
to calculate the noise power for the total radiation from the
cavity.
 Integrating the
expression (\ref{resolved}) over both frequencies yields the noise
power $S=\int d\omega d\omega' S_{\omega\omega'}$ for the total
photon flux leaving the system,
\begin{equation}
\label{totalnoise} S=\overline{J}
A^2\left(1+\frac{2\gamma}{w_0\Gamma\tau } \right) +\widetilde S
(1-A)^2,
\end{equation}
here $A=\int d\omega A_\omega$. By using the expressions
(\ref{n})-(\ref{phot}) for the mean distributions, the necessary
integrals are easily evaluated,
\begin{eqnarray*}
 \widetilde S &=&\frac{M(\Gamma^2-\gamma^2)}{32\gamma^7}
(5\Gamma^6-3\Gamma^4\gamma^2+7\Gamma^2\gamma^4+7\gamma^6),\\
A&=&\frac{M(\Gamma^4-\gamma^4)}{4PN_0\gamma^3 D},~~~
D=1+\frac{M(\Gamma^2-\gamma^2)^2}{4PN_0\gamma^3},
\end{eqnarray*}
Again, we analyze the obtained general expression neglecting
non-radiative losses, $\tau^{-1}=0$. By using the dimensionless
variables $p$ and $\eta$ defined in the preceding sections, the
Fano factor for the total photon flux leaving the cavity,
$F=S/\overline{J}$, can be written as, 
\begin{equation}
\label{totalfano}
F=\frac{4(1+g^2)^2+p^2(5-3g^2+7g^4+7g^6)}{4(1+g^2+2pg^3)^2}.
\end{equation}
The dimensionless width $g$ is a function of both the pumping
strength $p$ and the number of optic transitions $\eta$ and is
given by the solution of the equation (\ref{cubic}). The
dependence of the Fano factor (\ref{totalfano})  on the pumping
strength is plotted in Fig.\ 5. A surprising feature of this
dependence is the suppression of the noise {\it below Poissonian
level} for intermediate values of the pumping strength. This is
the demonstration of negative  correlations between the photons
due to their interaction with the gain media. Indeed, despite the
fact that the diagonal correlations (\ref{somega}-\ref{fomega})
increase monotonously with the pumping $p$ as a result of photon
bunching, the total flow of outgoing photons becomes less
fluctuating. For large values of $p$ the enhancement of the photon
bunching overwhelms the negative  correlations and the Fano factor
$F$ increases. Still, it remains strongly reduced compared to the
frequency-resolved Fano factor (\ref{fomega}), see Fig.\ 5.

 (i) For an open system, $\eta <1$, and large values of the
 pumping strength the Fano factor approaches the limit,
$$
F=\frac{1-2\eta+\frac{7}{4}\eta^2-\frac{7}{16}\eta^3}{(1-\eta)^3},
$$
and diverges at the critical point $\eta \to 1$. Its ratio,
$F/F_0$, to the frequency-resolved Fano factor (\ref{fomega1})
remains constant and decays monotonously from $1$ to $5/16$ while
$\eta$ changes from $0$ to $1$.

(ii)  Critical point, $\eta =1$. The Fano factor increases
quadratically for large pumping, $p \gg 1$,
$$
F=\frac{5 p^2}{36},
$$
with the ratio of the two factors remaining finite, $F/F_0 \to
5/36$.

(iii) Finally, in a  closed system, $\eta >1$, the fluctuations of
the total flux are strongly suppressed,
$$
F=\frac{5p^2}{4},
$$
compared to the diagonal fluctuations, with the ratio $F/F_0 \to
0$.

{\it Laser threshold}. So far the mean level spacing $\Delta$ has
not entered the expressions for the mean flux nor the noise power.
However, it sets the condition for applicability of the entire
approach. The above equations are semiclassical and  imply
incoherent radiation from the system. This is so provided that
there are many overlapping modes inside the cavity within the
linewidth $\gamma$. A characteristic mode width is given by the
inverse dwell time $\tau_{dw}^{-1}=\frac{M\Delta}{2\pi}$. As long
as $\gamma, \tau_{dw}^{-1} \gg \Delta$, the mode overlapping is
strong. Estimating the linewidth as $\gamma \approx \Gamma /\eta
p$ (for $\eta > 1$), we can establish the necessary conditions:
 $\frac{\Gamma^2 M}{PN_0} \gg \Delta,$ and $M \gg
1$.

Another condition for the semiclassical theory to be valid
requires the system to remain below the laser threshold. The
latter condition, for the $i$-th cavity mode, is determined by the
relation, $ \tau^{(i)}_{dw}/\tau^{(i)}_a \equiv s^{(i)} =1,$ with
the amplification time $\tau^{(i)}_a$ being in general a function
of the photon population of all cavity modes and, hence, some
complicated function of the pumping strength. From our equations
(\ref{photons}) and (\ref{n}) the semiclassical value of $s$ at
the center of the line $\omega=\omega_0$ is found,
$$
\overline{s}=\frac{\tau_{dw}}{a_{\omega_0}\overline{N}\Delta}=1-g^2<1,
$$
and the system stays below the threshold for any pumping
intensity. However, as the pumping increases, $g \to 0$, the
quantum distribution of $s^{(i)}$ around the mean value
$\overline{s}$ becomes important \cite{MiB}. At some critical
pumping, the mode with the largest $s^{(i)}$ reaches its threshold
value and starts to lase in a coherent fashion. This regime is
beyond the reach of the semiclassical theory. The problem of
fluctuations of radiation from a chaotic laser above the threshold
in the non-linear quantum regime remains an open one.

{\it Conclusions}. We presented a semiclassical theory of the
incoherent radiation from a chaotic cavity containing optically
active medium based on the radiative transfer theory with Langevin
sources. The theory reproduces the results of the random matrix
approach for fluctuations of radiation from a linear amplifier. By
accounting for the dynamics of the gain media the theory is
extended to incorporate the nonlinear effects: in particular, the
spectral line narrowing of the emerging light is calculated. The
fluctuations of radiation from the system are the result of two
competing effects: the enhancement of fluctuations due to photon
bunching  and the suppression of fluctuations due to interaction
with the gain medium.

I gratefully acknowledge fruitful discussions with C.\ Beenakker
and  B.\ Halperin. Conversations with A.\ Brataas and Ya.\
Tserkovnyak are also appreciated. This work was supported by the
National Science Foundation Grant DMR-02-33773 and by the Russian
Foundation for Basic Research Grant 01-02-16211.

\newpage

\begin{figure}[h]
\includegraphics[width=6.5cm]{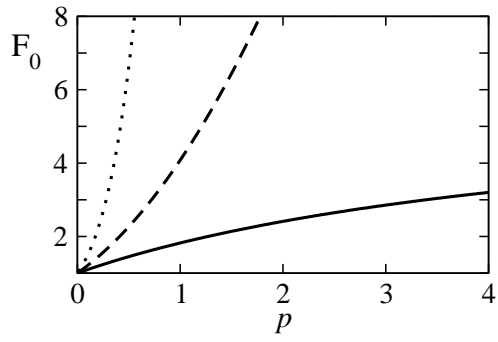}
\caption{The Fano factor for the fluctuations of radiation at the
center of the line, $\omega=\omega_0$, versus the dimensionless
pumping intensity (neglecting non-radiative losses) for different
values of the effective number of radiative transitions: $\eta=
1/2$ (solid line), $1$ (dashed line), $2$ (dotted line). }
\end{figure}

\begin{figure}[h]
\includegraphics[width=6.5cm]{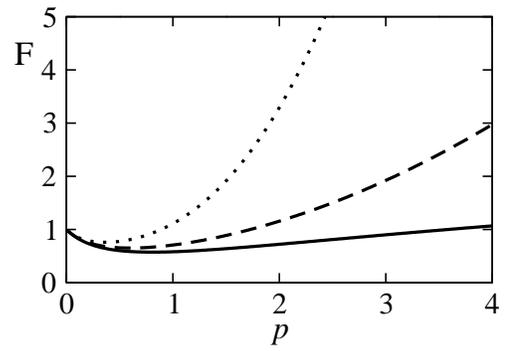}
\caption{The Fano facor for the total radiation from a chaotic
laser for different values of the effective number of radiative
transitions: $\eta= 1/2$ (solid line), $1$ (dashed line), $2$
(dotted line). }
\end{figure}

\newpage

\end{document}